\documentclass[conference]{IEEEtran}
\IEEEoverridecommandlockouts

\usepackage{cite}
\usepackage{amsmath,amssymb,amsfonts}
\usepackage{algorithmic}
\usepackage{graphicx}
\usepackage{textcomp}
\usepackage{xcolor}

\begin{document}

\title{In-depth Real-World Evaluation of NB-IoT\\ Module Energy Consumption}

\author{\IEEEauthorblockN{Milan Lukic, Srdjan Sobot, Ivan Mezei, Dejan Vukobratovic}
\IEEEauthorblockA{University of Novi Sad,\\ 
Faculty of Technical Sciences and Faculty of Sciences,\\
Trg Dositeja Obradovica 4/6, Novi Sad, Serbia\\
Emails: \{milan\_lukic, srdjansobot, imezei, dejanv\}@uns.ac.rs}
\and
\IEEEauthorblockN{Dragan Danilovic}
\IEEEauthorblockA{VIP Mobile\\
Bulevar Milutina Milankovica 1z,\\
Belgrade, Serbia\\
Emails: \{D.Danilovic\}@vipmobile.rs}
}

\maketitle

\begin{abstract}
Narrowband Internet of Things (NB-IoT) is a recent addition to the 3GPP standards offering low power wide area networking (LP-WAN) for a massive amount of IoT devices communicating at low data rates in the licensed bands. As the number of deployed NB-IoT devices worldwide increases, the question of energy-efficient real-world NB-IoT device and network operation that would maximize the device battery lifetime becomes crucial for NB-IoT service popularity and adoption. In this paper, we present a detailed energy consumption of an NB-IoT module obtained from a custom-designed NB-IoT platform from which a large amount of high time-resolution data is collected. We perform a detailed energy consumption analysis of each NB-IoT data transfer phase and discuss both the device and the network-side configurations that may affect the module energy consumption in each of the phases. Preliminary results and conclusions are put forth along with the discussion of ongoing and future study plans. 
\end{abstract}


\section{Introduction}
NB-IoT is the 3rd Generation Partnership Project (3GPP) standard for cellular Low-Power Wide Area Network (LP-WAN) and represents a forerunner of massive machine-type communications (mMTC) service envisaged in the upcoming 5G cellular networks \cite{Alvarino_2016, Wang_2017, Xu_2018}. After the initial tests and deployments were performed during 2017, and first commercial offerings presented in 2018, the NB-IoT services are entering the phase of widespread technology usage. The key question, both for the service users and for the mobile operators, is that of energy consumption of battery-operated NB-IoT devices. Indeed, understanding the details of NB-IoT device data transmission procedures, associated parameters that can be configured both at the device and at the network side, and their effects on the device energy consumption, is crucial for energy-efficient and scalable NB-IoT network operation.

The initial NB-IoT device energy consumption measurements and modelling studies are reported during the 3GPP standardization process \cite{3gpp_doc}. Motivated with the need to provide additional details, report user experiences, and present simpler and more accurate models, several recent studies focused on NB-IoT device energy consumption \cite{Lauridsen_2018, Martinez_2019, Maldonado_2019, Tsoukaneri_2020}. In the first study \cite{Lauridsen_2018}, the total device energy consumption of two commercial device platforms is measured in laboratory conditions, with the goal of proposing an energy consumption model for NB-IoT devices and estimating the battery lifetime. Martinez et al. \cite{Martinez_2019} provided energy consumption measurement study of two commercial NB-IoT platforms in real-world mobile operator network under different signal-to-noise (SNR) conditions. Detailed Markov Chain analysis and experimental validation in laboratory conditions of NB-IoT device consumption model is presented in \cite{Maldonado_2019}. Finally, in \cite{Tsoukaneri_2020}, detailed NB-IoT device energy consumption segmented across different data transmission phases is analyzed in laboratory conditions.

In this paper, we complement and expand the previous studies on NB-IoT device energy consumption. More precisely, the specific contributions of this study are:
\begin{itemize}
    \item We designed and fabricated a new NB-IoT device platform focusing on \emph{power consumption measurements of the NB-IoT module only}, thus making our study independent of the NB-IoT device platform.
    \item From our custom-designed platform, we collect fine time-resolution (ms-level) logs of: i) radio environment, ii) NB-IoT module power consumption, and iii) communication between the NB-IoT device and the base station.
    \item Similarly, but in more detail than in \cite{Tsoukaneri_2020}, we decompose and analyze the energy consumption across different data transmission phases, and provide precise phase-by-phase energy consumption measurements, which is conveniently visualised in this paper.
    \item Unlike previous studies, we point to the impact of downlink data transmission on energy consumption.
    \item We present experimental results emphasizing device and network-configurable parameters that affect the module's energy consumption in  each transmission phase.
\end{itemize}

\begin{figure*}
\centerline{\includegraphics[width=6.2in, height=2in]{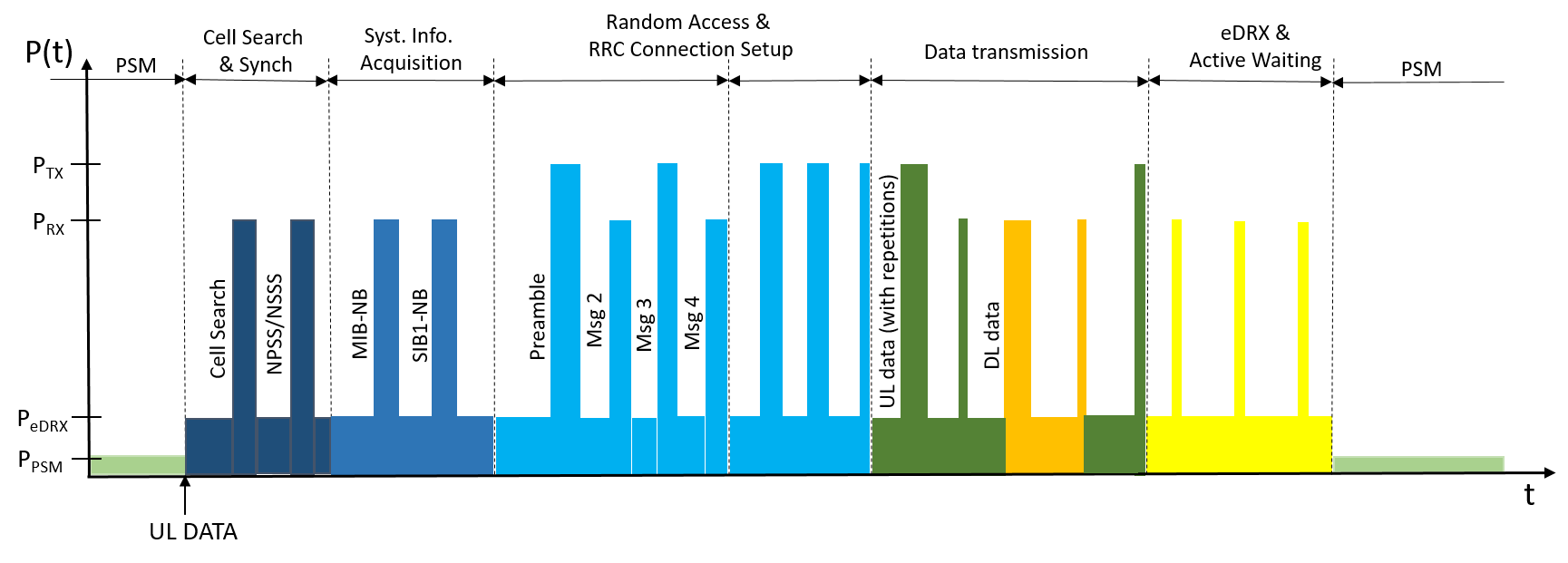}}
\caption{NB-IoT Uplink Transmission Procedure.}
\label{Fig_UL_Tx}
\end{figure*}

The paper is organized as follows. In Sec. II, we review in detail the NB-IoT uplink transmission procedures and 3GPP mechanisms for energy savings. We also present upper-layer options for uplink NB-IoT data transfer. Sec. III.A presents the details of our custom-designed NB-IoT platform used to generate data for our study. Visualisation of the collected data across different data transmission phases is presented via representative example in Sec. III.B. Sec. IV provides experimental results providing initial conclusions about influence of various device and network-side configurations on NB-IoT module energy consumption during each phase. Conclusions and plans for future work are given in Sec. \ref{section:conclusion}.

\section{Uplink Data Transfer in NB-IoT}

\subsection{NB-IoT Uplink Transmission Procedure}

In this section, we review the procedure of uplink (UL) data transmission, that repeats throughout the NB-IoT device lifetime. We note that similar descriptions with various level of details are available in \cite{Lauridsen_2018, Martinez_2019, Maldonado_2019, Tsoukaneri_2020, Maldonado_2017, Feltrin_2019}. 


From signaling perspective, in normal network operation, NB-IoT user equipment (UE) is attached to the network using Radio Resource Control (RRC) protocol. By default, UE spends time in RRC\_Idle state, in which the base station (eNB) cannot allocate resources to UE. If UE needs to send/receive data, it transits to RRC\_Connected state.

Fig. \ref{Fig_UL_Tx} shows the main phases in the UE originated UL data transmission. The UE has been in Power Saving Mode (PSM), however, the internal packet generation event triggered wake up and initiated UL transmission
comprising following phases:
\begin{itemize}
    \item \textbf{Cell Search and Synchronization:} UE wakes up from PSM mode to perform cell search and acquire downlink synchronization through NB-IoT synchronization signals. 
    \item \textbf{System Information Acquisition:} UE performs MIB/SIB system information blocks acquisition.
    \item \textbf{Random Access and RRC Connection Establishment:} UE initiates random access (RA) procedure by randomly selecting and transmitting RA preamble in the upcoming RA channel slot. If successful, the RA procedure involves exchange of four messages. RRC procedure of re-establishing RRC connection follows, finally establishing a data radio bearer.
    \item \textbf{UL/DL Data Transmission:} Once in RRC\_Connected state, a sequence of UL and DL transmissions follow, each preceded by eNB resource grants.
    \item \textbf{Active Waiting and eDRX:} Upon transmission, as configured by different timers, UE usually stays in active waiting until it transits to RRC\_Idle and eventually stays in eDRX mode until the timer expiration, after which it finally transits to PSM.
\end{itemize}

In Sec. III.B, we will revisit each stage in more detail through data transfer example log collected from our custom-designed NB-IoT platform described in Sec. III.A.

\subsection{NB-IoT Energy Savings Mechanisms}

The 3GPP introduced two key power saving features to optimize the UE power consumption: i) Power Saving Mode (PSM), and ii) extended Discontinuous Reception (eDRX) \cite{3gpp}. PSM enables the device to enter the state of minimal power consumption while being unreachable by the network. When in PSM, UE runs only Real Time Clock for keeping track of time, for the period set by Tracking Area Update (T3412) timer whose maximum value is approximately 413 days. eDRX allows the UE to stop monitoring radio channels and to enter low-power consumption mode for extended time period between paging occasions, dictated by Active Time value (T3324) of up to 186 minutes. Generally speaking, eDRX is more beneficial for devices that prefer a lower periodicity of reachability, where as PSM becomes an attractive alternative for devices that expect only infrequent mobile originating and terminating services.

If configured to use PSM, the UE exits PSM once it has UL data to transmit or is mandated to send periodic Tracking Area Update (TAU). During every attach and TAU procedure, UE requests Active Time T3324 value. If the network supports PSM and accepts that the UE uses PSM, it confirms by allocating T3324 to the UE. After UE completes data transmission, it first waits for the network to release the RRC connection to enter RRC\_Idle mode. This transition is controlled by a separate Inactivity Timer, of duration between 1 and 60 sec, which is configurable on the eNB side. The Inactivity Timer value is set as a trade off between signaling load due to frequent RRC connection setup requests and the UE power consumption.
A UE using PSM is available for mobile terminating services during the time of RRC\_Connected mode and the period of Active Time value after switching to the RRC\_Idle mode. After the Active Time value expires, the device moves to PSM. 

PSM is intended for UEs that are expecting only infrequent mobile originating and terminating services and that can accept a corresponding latency in the mobile terminating communication. In Attach and TAU procedures a PSM capable UE may request periodic T3412 timer value suitable for the latency of the service, thus controlling power consumption.

\subsection{Upper-Layer Protocols for NB-IoT Data Delivery}

\begin{table}[thbp]
\caption{Message overhead for various protocols.}
\begin{center}
\begin{tabular}{|c|c|c|c|c|}
\hline
 & IPv4 & UDP & MQTT-SN & Sum \\
Publish/ & 20 & 8 & 9 & 37 \\
\cline{2-5}
Subscribe & IPv4 & TCP & MQTT & Sum\\
 & 20 & 20 & 34 & 74 \\
\hline 
 & IPv4 & UDP & CoAP & Sum\\
Rest- & 20 & 8 & 20-30 & 48-58 \\
\cline{2-5}
based & IPv4 & TCP & HTTP & Sum\\
 & 20 & 20 & $>18$ & $>58$ \\
\hline
\end{tabular}
\label{table_1}
\end{center}
\end{table}

For end-to-end connectivity, upper layers (i.e. transport and application layers) target deployment of a reliable and efficient communication mechanism between an NB-IoT device and e.g. a cloud application. This mechanism can be implemented using IP-based or Non-IP data delivery. Non-IP solution is more bandwidth efficient as there is no overhead in addition to the payload data. However, chipset manufacturers encourage the usage of IP based data delivery in order to be network-agnostic. Non-IP data delivery is only possible in case the mobile network operator (MNO) has deployed an in-network platform server supported by NB-IoT chipset manufacturer which adds additional cost for MNOs and is chipset dependant. A truly scalable solution must rely on a network-agnostic protocol, where end devices interact directly with the cloud application residing outside the domain of the MNO.

For IP-based data delivery, NB-IoT equipment manufacturers support standard UDP and TCP transport layer protocols. Main advantage of UDP is that it is connection-less (i.e. no handshake is needed) while TCP is connection-oriented. Application layer protocols, handled by device firmware, need to be implemented in a way to reduce the bandwidth usage without compromising the quality of service. NB-IoT module providers provide standardized application programming interface (API) defined by the 3GPP as "AT command set for the user equipment" \cite{3GPP_AT}.

Two important classes of application layer protocols are REST-based and publish-subscribe protocols \cite{Al-Fuqaha_2015}. Although it is possible for developers to build a fully customized application layer protocol to suit the specific LP-WAN application requirements (e.g. in regards to bandwidth usage, latency, responsiveness, quality of service, security, and other criteria) this is usually not the case. A good solution favors lightweight protocols with minimum overhead. One such bandwidth-friendly publish-subscribe solution is MQTT-SN over UDP (instead of MQTT over TCP). CoAP over UDP instead of HTTP over TCP is an example of efficient REST-based solution. The solutions based on TCP are generally avoided due to several problems, such as higher overhead compared to UDP, maintenance of TCP connections, excessive delay in TCP acknowledgement deliveries, etc. Table I depicts the minimal amount of overhead to the payload (in bytes) for various protocols. In this paper we used plain UDP transmissions since it features the lowest overhead of only 28 bytes.

\section{NB-IoT Device Design, Data Collection and Visualisation}

\subsection{Custom-designed NB-IoT Device}
\label{section:nbiot_device}

\begin{figure}[ht]
\centerline{\includegraphics[width=3.5in]{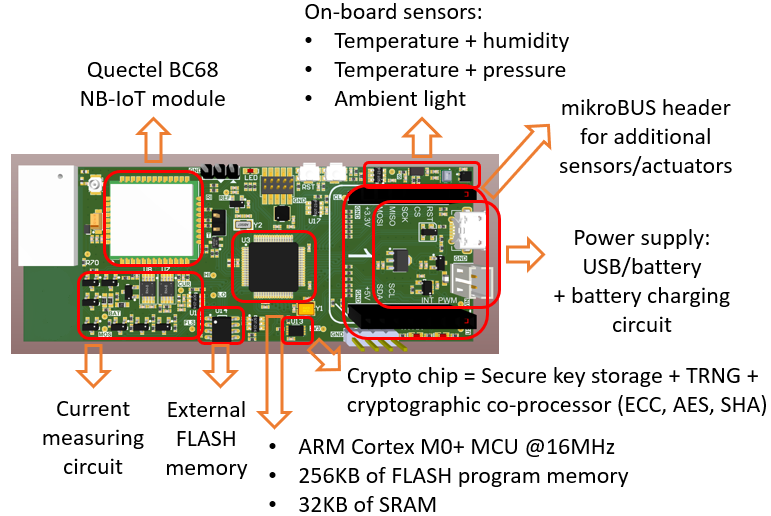}}
\caption{NB-IoT edge node.}
\label{Fig_Edge_node}
\end{figure}

For the purpose of building a testbed of NB-IoT devices within our research facilities, we have so far developed and tested a fully functional prototype of an edge node device. The device as shown in Fig. \ref{Fig_Edge_node} contains off-the-shelf NB-IoT module BC68 from Quectel.

The firmware inside on-board microcontroller initiates network registration and all subsequent communication tasks by issuing AT commands to the NB-IoT module via serial port. The additional serial port is available for monitoring log information, enabling us to track all interactions between NB-IoT protocol layers in real time, 
using UEMonitor (a log viewer tool). A current measuring circuitry is measuring the current consumption of the BC68 module in mA and uA range, which is convenient for power analysis both during periods of activity and in PSM. As opposed to experimental setup in other studies \cite{Lauridsen_2018, Maldonado_2019} which utilize external power measuring equipment, our on-board circuit will be a valuable asset for evaluation of power performance in scenarios where edge nodes are deployed en masse.

The overall hardware configuration of the edge node enables us to perform in-depth analysis, cross-referencing the information coming from three sources:
\begin{itemize}
    \item Radio channel conditions are available through AT commands, providing a snapshot of statistics for numerous parameters such as SNR, RSSI, total TX/RX time and power, BLER, etc. at a given moment in time when the command is executed. 
    \item NB-IoT module current consumption trace data. Our setup measures the current consumed by BC-68 NB-IoT module only, thereby eliminating the influence of other on-board components.
    \item Fine-grained protocol message exchange logs, available from UEMonitor application.
\end{itemize}

\begin{figure*}
\centerline{\includegraphics[width=7.5in, height=4in]{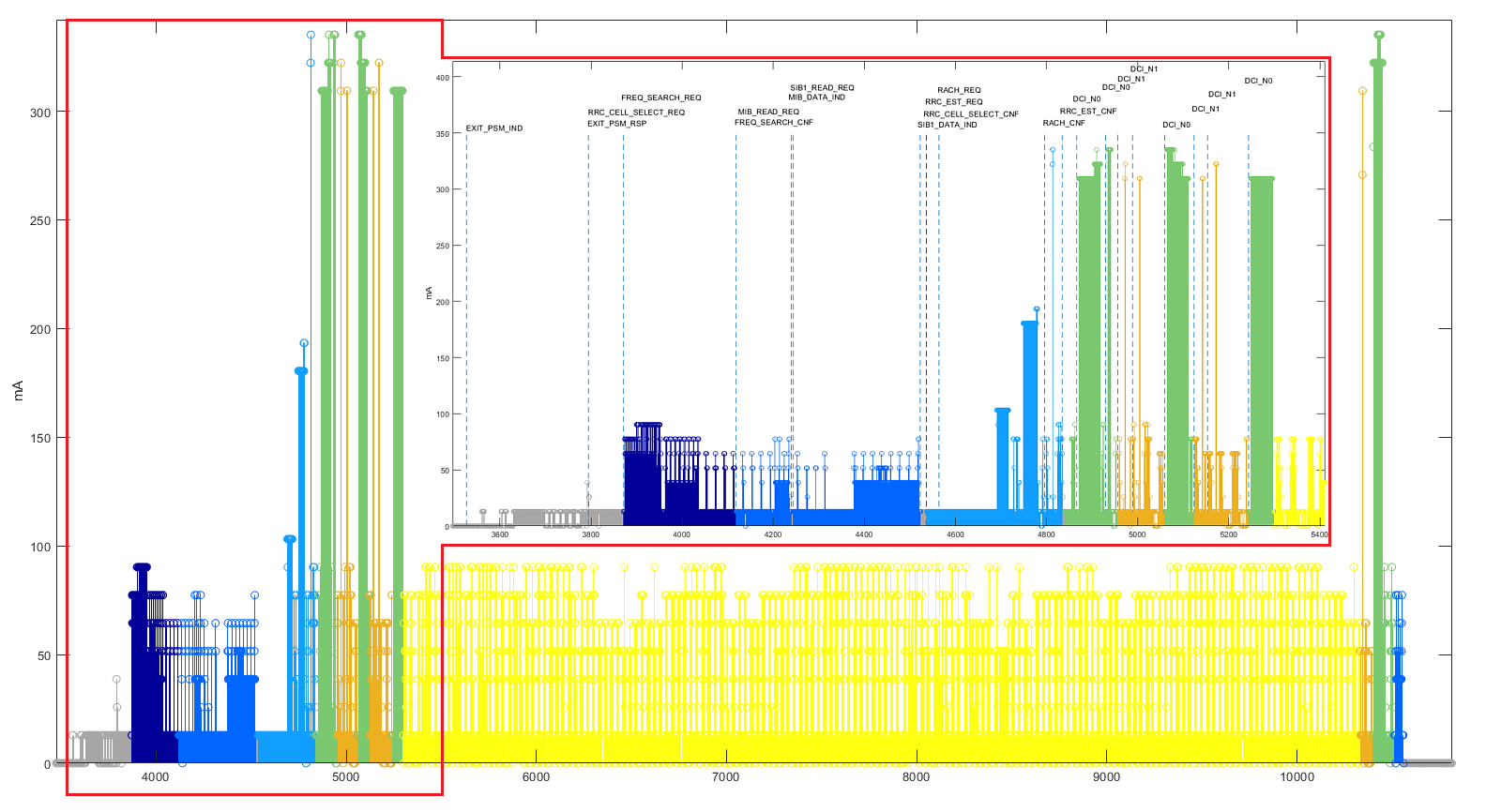}}
\caption{Detailed NB-IoT Uplink Transmission Procedure.}
\label{Fig_UL_Tx_detailed}
\end{figure*}

In the next subsection, we demonstrate and visualize an example of a post-processed log collected from our platform.

\subsection{Energy Costs of NB-IoT Uplink Data Transmission Phases}
\label{section:energy_costs}

The visualized output of the logs collected from our NB-IoT platform during the transmission of a single 64-bytes-long UDP packet are shown in Fig. \ref{Fig_UL_Tx_detailed}. The figure applies the same color code for different transmission phases as in Fig. \ref{Fig_UL_Tx}. For convenience, the initial part of transmission procedure (the part within the red rectangle) is presented in more detail, including explicit reference to messages exchanged between UE and the eNB and their precise timings, in "zoomed" area of Fig. \ref{Fig_UL_Tx_detailed}. 

\begin{figure}
\centerline{\includegraphics[width=3in]{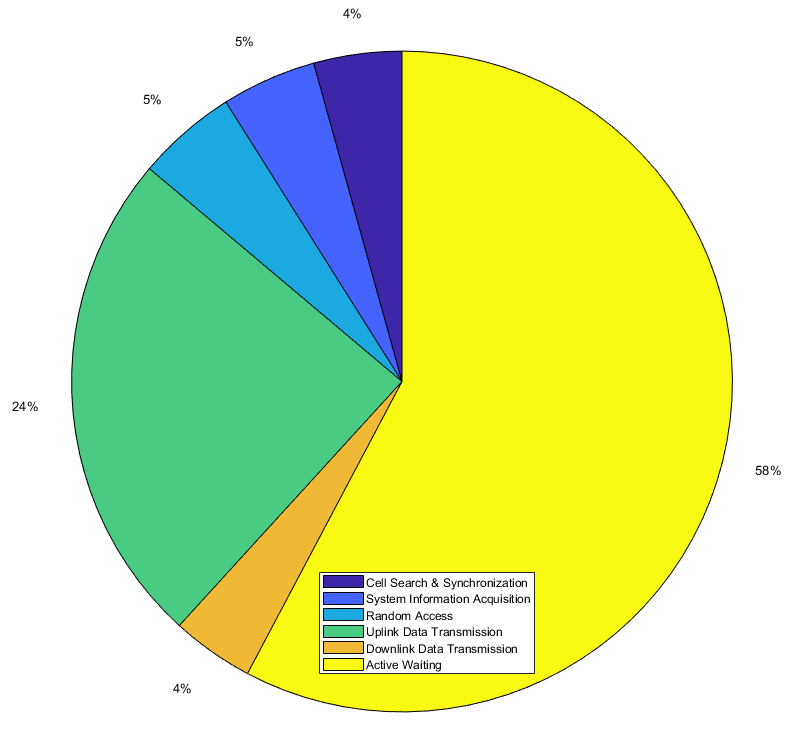}}
\caption{Energy consumption share of different phases during the transmission of a UDP packet.}
\label{Pie-chart}
\end{figure}

All the main phases of transmitting a UDP packet from the NB-IoT device to the base station begin with the appropriate request message (\emph{REQ}) and finish with the corresponding confirmation message (\emph{CNF}). Therefore, we are able to precisely measure the consumed energy of any phase. 

When a device wants to transmit data, it follows the connection establishment procedure. After a full frequency scan to find a suitable NB-IoT cell, downlink Narrowband Primary Synchronization Signal (NPSS) is used to achieve synchronization in time and frequency and Narrowband Secondary Synchronization Signal (NSSS) is used to detect the cell identity number. In our example, the aforementioned \emph{Cell Search and Synchronization} procedure has happened between \emph{FREQ\_SEARCH\_REQ} and \emph{FREQ\_SEARCH\_CNF} messages. Hence, all the samples between these two messages represent energy consumption for cell search and synchronization phase. 

After the device has performed synchronization, it starts with the \emph{System Information Acquisition} phase to obtain the master and system information blocks (MIB-NB and SIB1-NB) necessary to complete cell selection. In our example in Fig. \ref{Fig_UL_Tx_detailed}, approx. 10\% of total energy consumption is used for the first two procedures. This consumption will depend on the reception conditions, device and base station parameters, which we observe in more detail in Sec. IV. 

Next, the \emph{Random Access} (RA) phase follows with standard preamble-based four-message exchange, where, in our example, another 6\% of total energy consumption is used. 

After the attach process, the device continues to \emph{Data Transmission} phase. Downlink control information (DCI) is utilized to indicate for reception and transmission of data, containing uplink scheduling grants and downlink scheduling assignments. Useful information on transport block sizes, modulation and coding configuration and the number of repetitions are part of DCI messages. NB-IoT has three DCI formats: i) N0: used for UL grants, ii) N1: used for DL scheduling, and iii) N2: used for paging. In this example, around 20\% of total energy consumption is used by uplink data transmission and only 1\% for downlink data reception. 

After the data transmission and reception, the device remains connected to the network for 5 seconds, in the so-called \emph{Active waiting} phase controlled by Inactivity Timer. This process consumes around 64\% of total energy, making it by far the most energy-consuming process \cite{Tsoukaneri_2020}. After 5 seconds, release of RRC connection is done and, after short eDRX phase controlled by Active Time value (T3342) timer, the device enters the PSM. The total energy consumption disaggregated over different phases is illustrated in Fig \ref{Pie-chart}.

\section{Experimental Results}

\begin{figure}
\centerline{\includegraphics[width=2in]{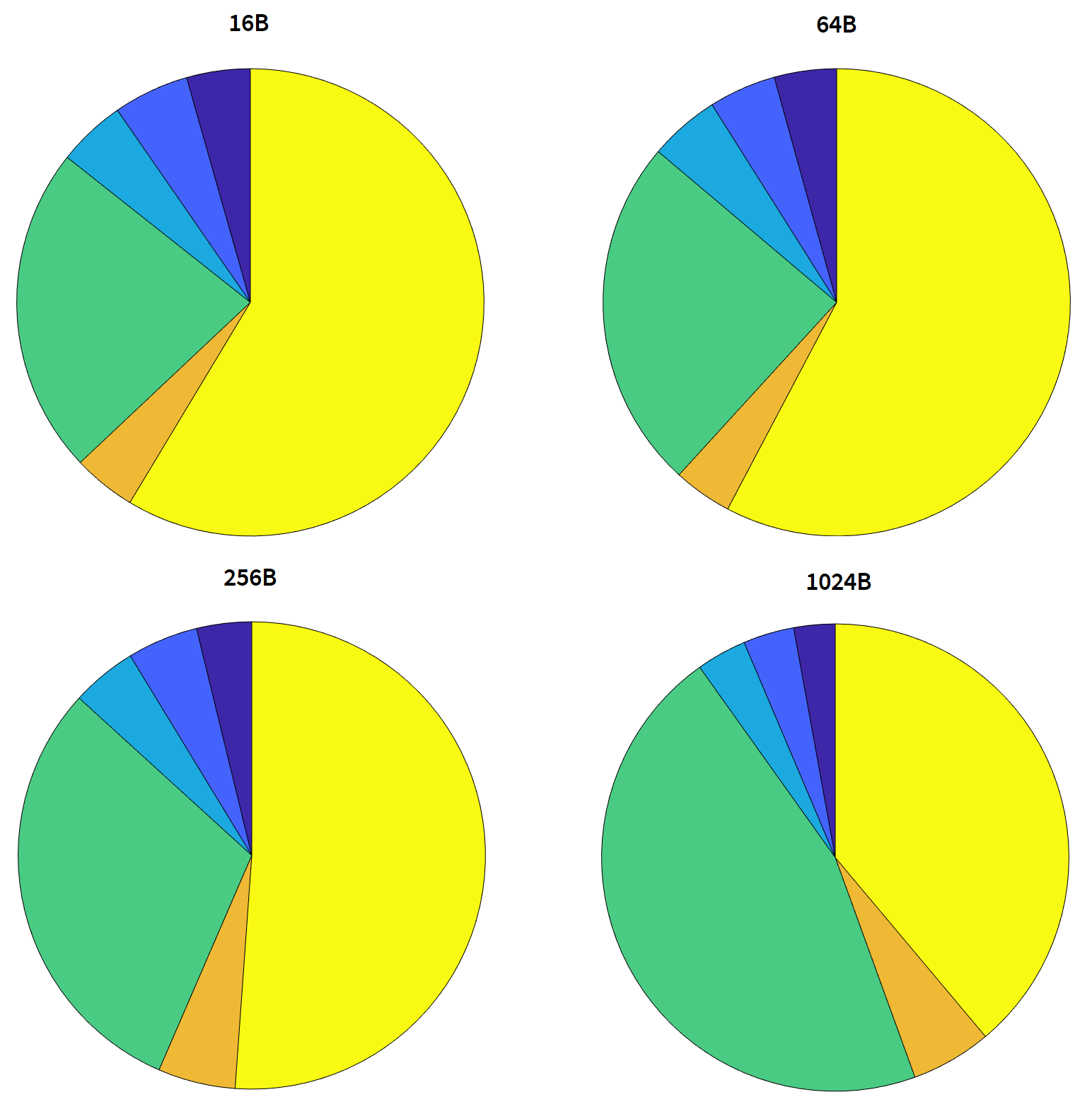}}
\caption{Energy consumption share of different phases during the transmission of 16, 64, 256, and 1024-bytes-long UDP packet}
\label{Pie-chart-different-sizes}
\end{figure}

\begin{figure}[htpb]
\centerline{\includegraphics[width=3.3in]{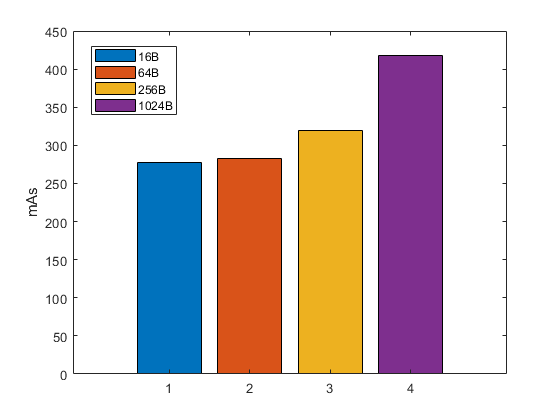}}
\caption{Total absolute energy consumption in [mAs] of the transmission of 16, 64, 256, and 1024-bytes-long UDP packet}
\label{mAs-total}
\end{figure}

\begin{figure}
\centerline{\includegraphics[width=2in]{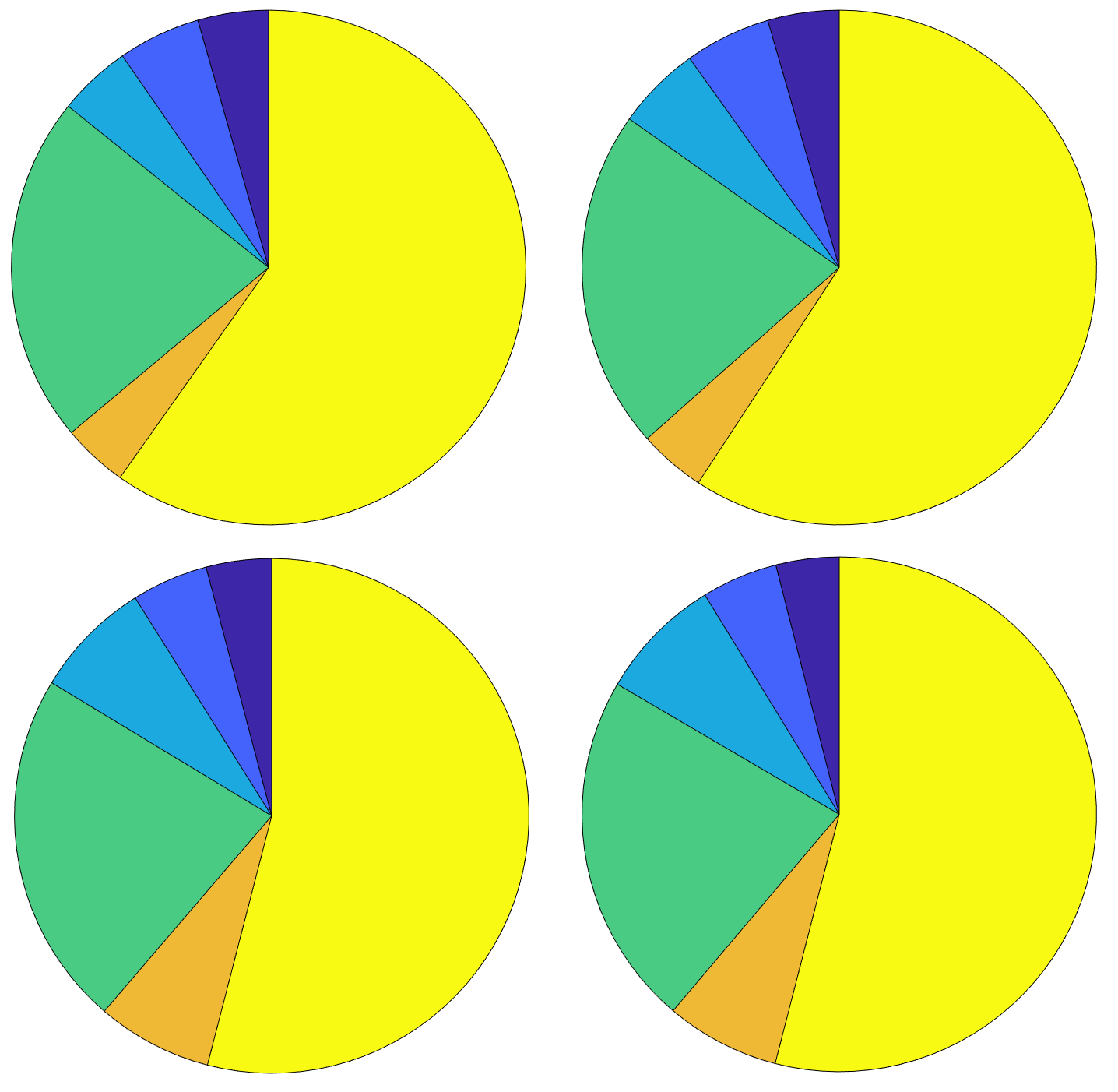}}
\caption{Energy consumption share of different phases during the transmission of a 64-bytes-long UDP packet in different radio conditions.}
\label{Pie-chart-same-size-diff-loc}
\end{figure}

All the experiments and measurements have been conducted using our test device, as described in section \ref{section:nbiot_device}. The device was connected to a PC through 2 serial ports. The first one is BC68 debug port, allowing us to log NB-IoT protocol messages using UEMonitor application. The other serial port was used to initiate transmissions from the device, and to to receive data from the device, including current samples (sampling period = 1ms), and statistics from BC68 module.

In the first experiment, we tested bidirectional communication between our device and remote server, using UDP packets of 4 different sizes: 16B, 64B, 256B and 1024B. Upon the reception of an uplink UDP packet, the server immediately returns the echo, i.e. it sends back the same content in a downlink UDP packet. We were sending different sizes of a UDP packet to the base station from the same location, thus radio conditions were unchanged. As one can see in Fig \ref{Pie-chart-different-sizes}, the total energy consumption share of Data Transmission phase increases with the size of the packet and eventually overtakes the consumption of Active Waiting phase set to 5s period. Apart from the two phases, the other phases remain approximately the same, as the reception conditions remain the same. We complete this example by illustrating in Fig. \ref{mAs-total} the total energy consumption of UDP packet transmission of varying lengths ranging from 16B to 1024B.

In the second experiment, we fixed the size of a packet to 64 B but varied the radio conditions. Fig \ref{Pie-chart-same-size-diff-loc} shows the energy consumption share of different phases from four separate locations around the base station, ranging from about 100m distance from eNB (the center of the cell: the upper left corner) up to about 1000m away from eNB (the edge of the cell: the lower right corner). We monitored both the radio parameters and examined DCI N0 and DCI N1 messages. In the UL, we could not spot significant changes in the performance of Uplink Data Transmission phase, where the number of packet repetitions remain at minimum, i.e., equal to one (possibly due to uplink power control, however, without significant impact on device power consumption). Unlike the uplink, the energy consumption share of Downlink Data Transmission phase (orange colour) was gradually increasing towards the cell edge, as shown in Fig \ref{Pie-chart-same-size-diff-loc}. This is mainly due to the decrease in received signal-to-noise ratio (SNR), and consequently, the number of packet repetitions gradually increased from one up to four repetitions per packet\footnote{Note that the testing is done in urban scenario with the cell size limited to about 1000 m due to neighbouring cells.}.

\begin{figure}
\centerline{\includegraphics[width=2in]{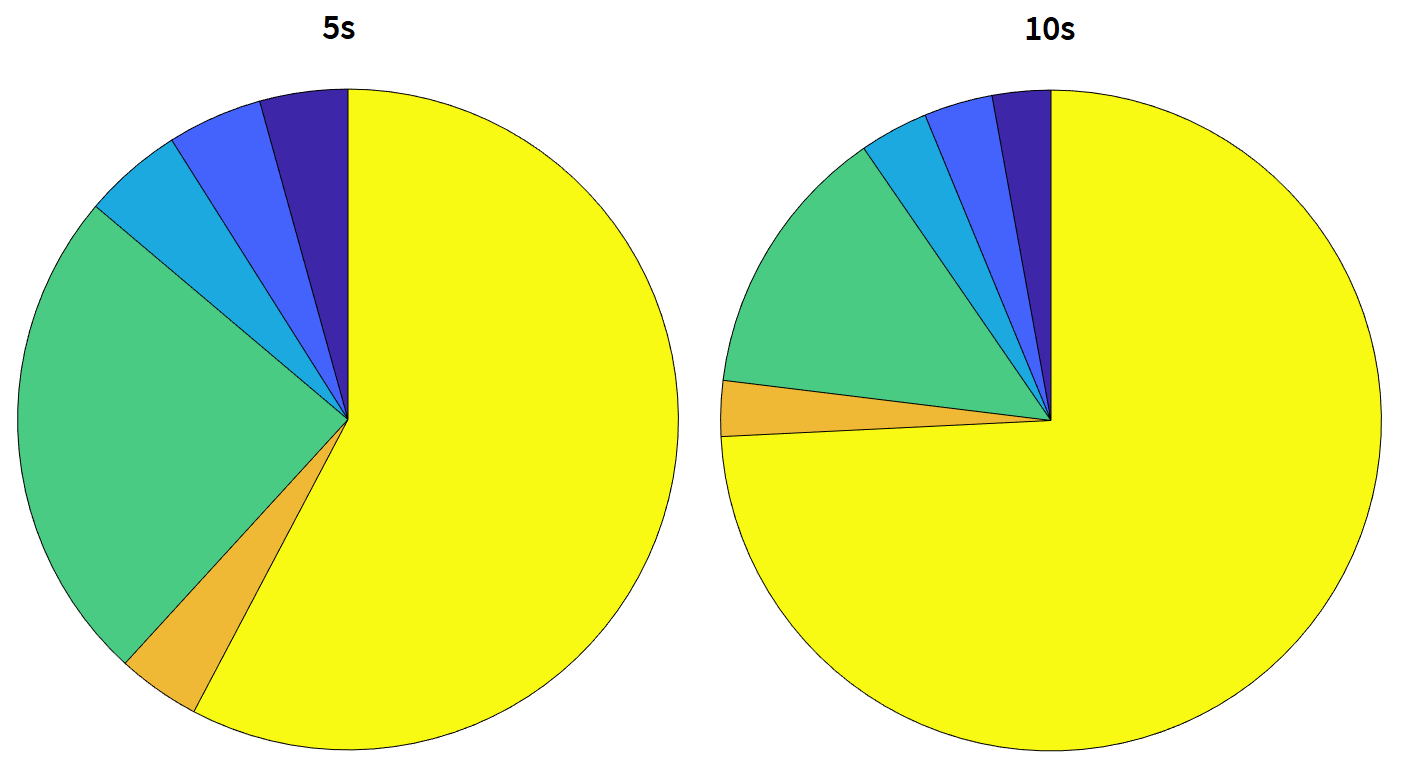}}
\caption{Energy consumption share of different phases during the transmission of 64-bytes-long UDP packet, with Inactivity Timer set to 5 \& 10 sec.}
\label{pie_10_sec_active}
\end{figure}

In the third experiment, we transmitted UDP packet of length 64B in identical radio conditions, but with extended Inactivity Timer set to 10 seconds. In all previous examples, we used the default value of Inactivity timer of 5 seconds set at the eNB. Fig. \ref{pie_10_sec_active} clearly shows strong impact of Inactivity Timer setting, where Active Waiting phase for the 10s setup consumes almost 3/4 of the total energy spent on transmitting 64B UDP packet. Clearly, as noted in \cite{Tsoukaneri_2020}, one needs to be careful when and to how large value to set the Inactivity Timer, as it will affect per-packet energy consumption.

For some applications, Inactivity Timer could result in high energy waste. In particular, for applications that do not require IoT device to wait for additional downlink data. In such cases, UE should switch to PSM state as soon as possible after receiving expected data. In order to allow faster transition to PSM state, in Release 13, for Control Plane optimisation, \emph{Release Assistance} feature is introduced that allows the UE to request for the RRC connection release as soon as the message has been received by the network \cite{Release-Assistance}. In such case, the MME can immediately trigger the Release procedure reducing the UE time spent in RRC\_Connected state waiting for additional transmissions.

\begin{figure}
\centerline{\includegraphics[width=2in]{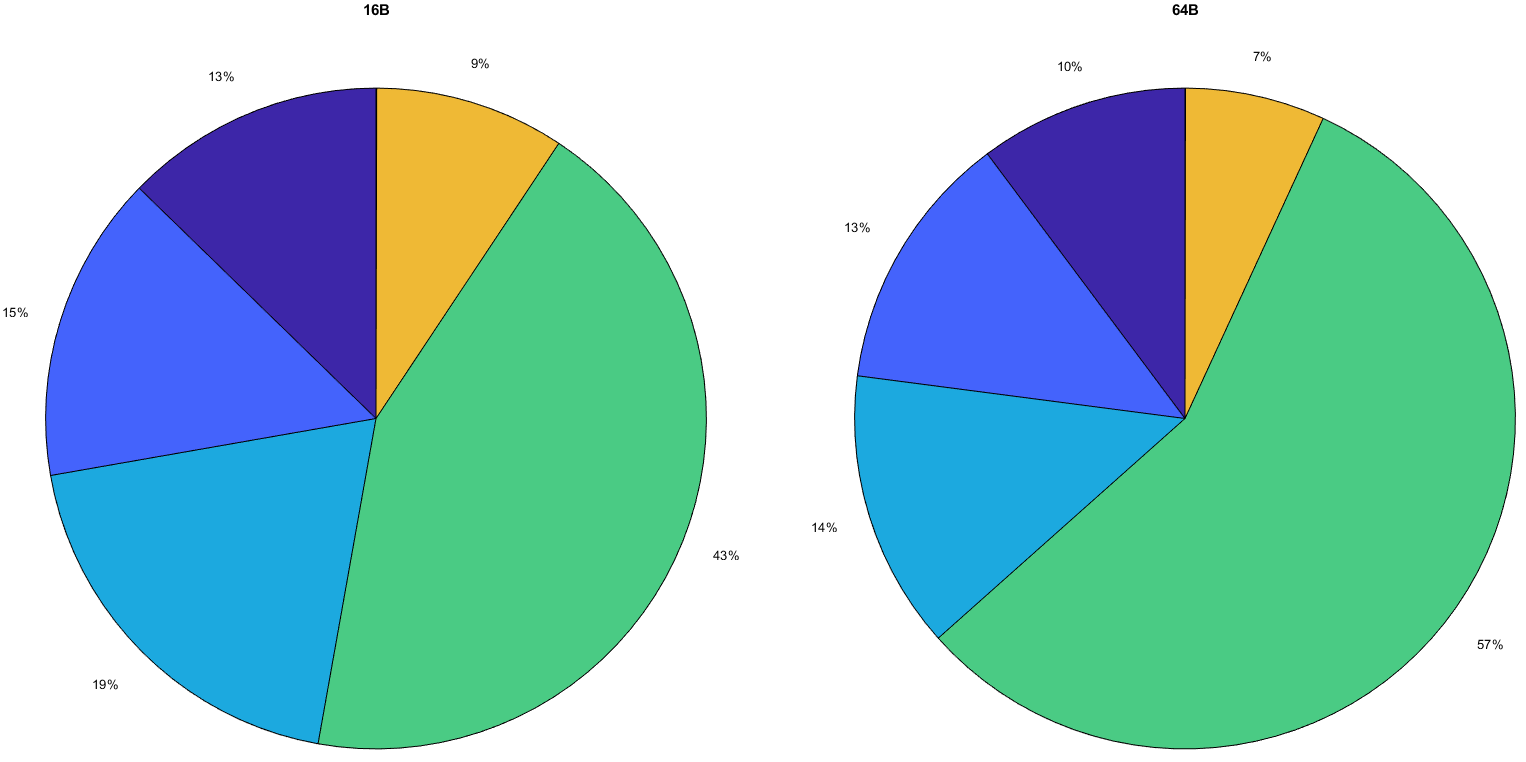}}
\caption{Energy consumption share of different phases during the transmission of 16 and 64-bytes-long UDP packet with Release Assistance.}
\label{pie_11_rel_ass}
\end{figure}

Fig. \ref{pie_11_rel_ass} illustrates the energy consumption of the experiment in which UDP packets of lengths 16B and 64B are sent using the Release Assistance feature. One can clearly note that Active Waiting (yellow coloured) phase is now eliminated, resulting with extremely energy efficient one-shot transmission of a single UDP packet. 

\section{Conclusion}
\label{section:conclusion}

In this paper, we presented our custom-designed NB-IoT platform for fine-grained measurements of energy consumption of an NB-IoT module across different phases during the data transmission. Using the designed platform in a real-world setup supported by a mobile operator, we presented several numerical examples of NB-IoT module energy consumption under various settings. As our future work, we will increase the scale of our experiments by connecting about one hundred of NB-IoT nodes to an eNB, while collecting large energy consumption data sets under various parameter configurations, both at the UE and at the eNB side.

\section*{Acknowledgment}

This work has received funding from the European Union Horizon 2020 research and innovation programme under the grant agreements No. 833828 and No. 856697.

\end{document}